\documentclass[12pt]{iopart}
\usepackage{iopams}
\usepackage{graphicx}
\usepackage{subcaption}
\usepackage{hyperref}
\usepackage{hypernat}
\usepackage{color}
\usepackage{amssymb}
\usepackage{bbm,bbold}
\usepackage{cite}

\begin{document}
\title[Mean first-encounter times of random walkers with resetting on networks]{Mean first-encounter times of simultaneous random walkers with resetting on networks}
	\author{Daniel Rubio-G\'omez$^1$, Alejandro P. Riascos$^2$\footnote{E-mail Corresponding Author: alperezri@unal.edu.co}, and Jos\'e L. Mateos${}^{1,3}$}  
	\address{${}^1$Instituto de F\'isica, Universidad Nacional Aut\'onoma de M\'exico, 
		C.P. 04510, Ciudad de M\'exico, M\'exico\\
		${}^2$ Departamento de Física, Universidad Nacional de Colombia, Bogotá, Colombia\\
		${}^3$ Centro de Ciencias de la Complejidad, Universidad Nacional Aut\'onoma de M\'exico, C.P. 04510, Ciudad de M\'exico, M\'exico}
\date{\today}
\begin{abstract}
We investigate the dynamics of simultaneous random walkers with resetting on networks and derive exact analytical expressions for the mean first-encounter times of Markovian random walkers. Specifically, we consider two cases for the simultaneous dynamics of two random walkers on networks: when only one walker resets to the initial node, and when both walkers return to their initial positions. In both cases, the encounter times are expressed in terms of the eigenvalues and eigenvectors of the transition matrix of the normal random walk, providing a spectral interpretation of the impact of resetting. We validate our approach through examples on rings, Cayley trees, and random networks generated using the Erdős–Rényi, Watts–Strogatz, and Barabási–Albert algorithms, where resetting significantly reduces encounter times. The proposed framework can be extended to other types of random walk dynamics, transport processes, or multiple-walker scenarios, with potential applications in human mobility, epidemic spreading, and search strategies in complex systems.
\end{abstract}


\maketitle
\section{Introduction}
Diffusive transport and random walk strategies have been extensively employed across diverse disciplines as effective mechanisms for locating hidden targets or exploring specific regions in space. Notable applications include animal foraging behavior \cite{ViswaBook2011}, the dynamics of urban transportation systems \cite{LoaizaMonsalvePlosOne2019,RiascosMateosSciRep2020}, protein searches for binding sites along DNA strands \cite{Coppey2004}, and information retrieval and ranking in databases \cite{LeskovecBook2014,BlanchardBook2011}, among many others. In this broad context, recent years have witnessed growing interest in stochastic search processes incorporating resetting or restart mechanisms. When a random process is intermittently interrupted and restarted from a predefined configuration, typically its initial state, the statistical properties of the dynamics are markedly modified. Remarkably, the average time required to reach a specified target, known as the mean first-passage time, can often be minimized by tuning the resetting rate \cite{evans2011diffusion,Evans2011JPhysA,reuveni2016optimal,EvansReview2019}.
Various resetting protocols have been proposed and analyzed \cite{pal2016diffusion,nagar2016diffusion,Bhat2016JStat,chechkin2018random,PalPRE_2019,Nagar_2023,SalgadoPRE_2024}, applied to a range of underlying stochastic processes, such as Brownian motion \cite{evans2011diffusion,Evans2011JPhysA,MajumdarPRE2015}, biased diffusive dynamics \cite{montero2013monotonic,ray2019peclet}, and anomalous diffusion models \cite{Kusmierz2014PRL,kusmierz2015optimal,kusmierz2019subdiffusive,maso2019transport}.
\\[2mm]
Moreover, a wide range of phenomena can be effectively modeled as dynamical processes on networks \cite{NewmanBook,barabasi2016book,VespiBook}. The interplay between network topology and the dynamics taking place is central to understanding the behavior of many complex systems \cite{NewmanBook,VespiBook,VanMieghem2011}. In this context, random walk strategies, where transitions occur only between adjacent nodes (normal random walk), provide a natural and widely applicable framework for analyzing diffusive transport on networks \cite{VespiBook,Hughes,Lovasz1996,MulkenPR502}. Significant progress has been made in understanding network exploration via random walks \cite{NohRieger2004,Tejedor2009PRE,MasudaPhysRep2017}, including extensions to non-local strategies that incorporate long-range transitions between distant nodes \cite{RiascosMateos2012,RiascosMateosFD2014,Weng2015,Guo2016,Michelitsch2017PhysA,deNigris2017,Estrada2017Multihopper}, as well as the study of collective behavior involving multiple simultaneous walkers \cite{WengPRE2017,WengChaos2018,AgliariPRE2016,AgliariPRE2019,Riascos_Multiple2020}. Despite this progress, random walks on networks subject to resetting have received comparatively less attention \cite{Avrachenkov2014,Avrachenkov2018,Touchette_PRE2018,ResetNetworks_PRE2020,christophorov2020peculiarities,Wald_PRE2021,bonomo2021first}. Recent advances have established connections between random walk dynamics with resetting to a single node and the spectral decomposition of the transition matrix that defines the normal random walk with local hops between neighboring nodes with equal probability \cite{Touchette_PRE2018,ResetNetworks_PRE2020,MultipleResetPRE_2021}. These insights underscore the potential of resetting mechanisms as efficient strategies for exploring diverse network topologies \cite{ResetNetworks_PRE2020,MultipleResetPRE_2021,Riascos_JPA_2022,Michelitsch2025}.
\\[2mm]
\begin{figure}[!t]
	\begin{center}
		\includegraphics*[width=0.7\textwidth]{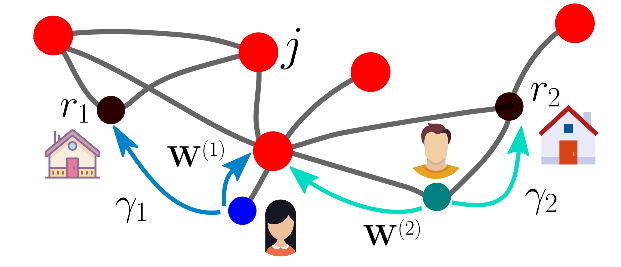}
	\end{center}
	\vspace{-5mm}
	\caption{\label{Fig_1} Two random walkers move on a street network where nodes represent locations. Alice moves between nodes according to a transition matrix $\mathbf{W}^{(\mathrm{1})}$, while Bob follows a different strategy defined by $\mathbf{W}^{(\mathrm{2})}$. Additionally, both walkers have a probability $\gamma_1$ and $\gamma_2$ of returning to their respective homes at nodes $r_1$ and $r_2$.}
\end{figure}
Most of the aforementioned studies focus on the dynamics of a single random walker, while the simultaneous motion of multiple walkers has received comparatively limited attention \cite{WengPRE2017}. Nevertheless, the presence of multiple agents is a common feature in real-world processes occurring on complex systems; for instance, in encounter networks arising from human activity \cite{RiascosMateosPlosOne2017,Mastrandrea_PlosOne2015}, epidemic spreading \cite{SatorrasPRL2001,ValdezBraunsteinHavlin2020,Bestehorn2021,Granger2024}, ecological interactions \cite{GiuggioliPRL2013,CraftRoyalB_2015}, and the emergence of extreme events \cite{KishorePRL2011}, among others. Recent efforts examining the efficiency of multiple searchers in locating a target on networks include the study of the mean time required for one or more walkers to reach the target \cite{DaiPhysA2020}, the emergence of universal laws governing search times \cite{WengPRE2017,WengPRE2018universal,DaiPhysA2020}, analytical descriptions of encounter times among many random walkers \cite{DPSandersPRE2009,Riascos_Multiple2020}, and search strategies involving the capture of moving targets with predefined trajectories \cite{Weng_2017,WengChaos2018}. Moreover, other recent studies underscore the potential of collective resetting mechanisms in systems with multiple agents, where simultaneous resets~\cite{BiroliPRL_2023,BiroliPRE_2023} or threshold-triggered restarts~\cite{biswas2025_arxiv} induce long-range correlations that enhance performance in optimization and search tasks.
\\[2mm]
In this paper, we develop a theoretical framework to investigate the collective dynamics of simultaneous, non-interacting Markovian random walkers with resetting, where each walker evolves according to its own transition matrix. For the case in which the transition matrices are diagonalizable, we derive exact analytical expressions for the mean first-encounter times. Figure \ref{Fig_1} illustrates this process for two agents navigating a network by visiting nodes connected by edges. At each time step, each walker can reset to a designated node, $r_1$ or $r_2$, with probabilities $\gamma_1$ and $\gamma_2$, respectively. With complementary probabilities $1 - \gamma_1$ and $1 - \gamma_2$, they follow a random walk strategy with transition probability matrices  $\mathbf{W}^{(\mathrm{1})}$ and  $\mathbf{W}^{(\mathrm{2})}$. Although the walkers do not interact directly, it is of central interest to determine whether and when they \emph{encounter} each other; that is, occupy the same node simultaneously. We are interested in the effect of stochastic reset in the mean number of steps required for such encounters to occur for the first time at a given  node $j$.
\\[2mm]
The paper is organized as follows: In Sec. \ref{Sec_GeneralTheory}, we present the general theoretical framework, which includes the analysis of a single walker under resetting as well as the formulation for simultaneous non-interacting random walks on networks. Building upon this formalism, in Sec. \ref{Sec_Simul_RW_reset} we derive analytical expressions characterizing the mean first-encounter times of multiple walkers on networks, where one or more of the walkers follow a resetting strategy. In particular, we obtain two analytical expressions for the mean first-encounter time of two walkers. The first result corresponds to the case where only one of the walkers resets to its initial position, while the second addresses the scenario in which both walkers restart to the initial node. In both cases, the encounter times are expressed in terms of the eigenvalues and eigenvectors of the transition matrix that defines the normal random walk without resetting. We explore a variety of illustrative examples to analyze the effect of resetting across different network topologies, including rings, Cayley trees, and random networks generated via the Erdős–Rényi, Watts–Strogatz, and Barabási–Albert models. Our findings show different cases where the resetting dynamics significantly reduce the mean first-encounter times of the random walkers. In Sec. \ref{Sec_Conclusions}, we present the conclusions. 
The formalism introduced in this work can be extended to the study of simultaneous processes with resetting involving other types of random walkers and diffusive transport dynamics on networks.
\section{General theory}
\label{Sec_GeneralTheory}
\subsection{Random walks with reset on networks}
Let us consider a random walker on an arbitrary connected network with $N$ nodes, labeled as $i = 1, \dots, N$. We study the discrete-time evolution of a random walker, which starts at node $i$ at time $t = 0$. At each time step, the walker follows one of two possible actions: with probability $1 - \gamma$, it performs a random jump to a different node of the network, while with probability $\gamma$, it resets to a fixed node $r$. In the absence of resetting ($\gamma = 0$), the probability of hopping from node $l$ to node $m$ is given by $w_{l\to m}$, and we assume that the random walk is ergodic and governed by the transition matrix $\mathbf{W}$, whose elements $w_{l\to m}$ define the transition probabilities for $l, m = 1, \dots, N$. The transition matrix may represent either local transitions, i.e., hops between directly connected nodes, or non-local transitions, allowing hops between distant nodes. However, for the application of our approach, it is necessary to consider only transition matrices $\mathbf{W}$ that are diagonalizable (see Ref. \cite{reviewjcn_2021} for examples of random walk strategies where this condition is satisfied). The occupation probability of the process under resetting follows the master equation \cite{ResetNetworks_PRE2020}
\begin{equation}
	\label{mastereq1}
	P_{ij}(t+1;r,\gamma) = (1-\gamma)\sum_{l=1}^N P_{il}(t;r,\gamma)w_{l\to j}+\gamma\delta_{rj},
\end{equation}
here $P_{ij}(t;r,\gamma)$ denotes the probability to find the walker at node $j$ at time $t$, given the initial position $i$, resetting node $r$ and resetting probability $\gamma$ ($\delta_{rj}$ denotes the
Kronecker delta). The first term in the right-hand side of Eq. (\ref{mastereq1}) represents hops associated to the transition probabilities  $\mathbf{W}$ and the second term describes resetting to $r$. With the introduction of the transition probability matrix $\mathbf{\Pi}(r;\gamma)$ with elements $
\pi_{l \to m}(r;\gamma)\equiv (1-\gamma) w_{l\to m}+\gamma\,\delta_{rm}$, Eq. (\ref{mastereq1}) takes the simpler form \cite{ResetNetworks_PRE2020}
\begin{equation}\label{mastermarkov}
	P_{ij}(t+1;r,\gamma) = \sum_{l=1}^N  P_{il} (t;r,\gamma) \pi_{l\to j}(r;\gamma),
\end{equation}
where $\sum_{m=1}^N \pi_{l \to m}(r;\gamma)=1$. The matrix $\mathbf{\Pi}(r;\gamma)$ fully characterizes the resetting process, ensuring that all nodes of the network remain accessible as long as the resetting probability satisfies $0\leq \gamma < 1$. Both $\mathbf{W}$ and $\mathbf{\Pi}(r;\gamma)$ are stochastic matrices, whose eigenvalues and eigenvectors provide key insights into the system's dynamics. In particular, knowing these spectral properties allows the calculation of the occupation probability at any time, including the stationary distribution at $t \to \infty$, as well as the mean first-passage time to any node \cite{ResetNetworks_PRE2020,MultipleResetPRE_2021,Riascos_JPA_2022}. 
\\[2mm]
In the following, we use Dirac's notation for the eigenvectors. We denote the eigenvalues of the matrix $\mathbf{W}$ as $\lambda_l$, where $\lambda_1 = 1$, and its right and left eigenvectors as $\left|\phi_l\right\rangle$ and $\left\langle\bar{\phi}_l\right|$, respectively, for $l = 1,2,\ldots,N$.  These eigenvectors form an orthonormal basis and satisfy the relations  $\left\langle\bar{\phi}_l|\phi_m\right\rangle=\delta_{lm}$ and $\sum_{l=1}^N \left|\phi_l\right\rangle \left\langle\bar{\phi}_l\right|=\mathbb{1}$, where $\mathbb{1}$ is the $N\times N$ identity matrix.  
Similarly, the eigenvalues of $\mathbf{\Pi}(r;\gamma)$ are denoted as $\zeta_l(\gamma)$, with corresponding right and left eigenvectors $\left|\psi_l(r;\gamma)\right\rangle$ and $\left\langle\bar{\psi}_l(r;\gamma)\right|$, respectively \cite{ResetNetworks_PRE2020}.
\\[2mm]
Moreover, the eigenvalues and eigenvectors of $\mathbf{\Pi}(r;\gamma)$ are directly related to those of $\mathbf{W}$, which is recovered in the limit $\gamma \to 0$ \cite{ResetNetworks_PRE2020}.
The connection between the eigenvalues $\lambda_l$ and $\zeta_l(\gamma)$ is obtained from the relation 
\begin{equation}\label{Def_MatPi_R1}
	\mathbf{\Pi}(r;\gamma)=(1-\gamma)\mathbf{W}+\gamma \mathbf{\Theta}(r),
\end{equation}
where the elements of the matrix $\mathbf{\Theta}(r)$ are $\Theta_{lm}(r)=\delta_{mr}$. Namely, $\mathbf{\Theta}(r)$ has entries $1$ in the $r^{th}$-column and null entries everywhere else, therefore (see Ref. \cite{ResetNetworks_PRE2020} for details)
\begin{equation}\label{eigvals_zeta}
	\zeta_l(\gamma) = \left\{
	\begin{array}{ll}
		1 \qquad & \mathrm{for}\qquad l=1, \\
		(1-\gamma)\lambda_l \qquad & \mathrm{for}\qquad l=2,3,\ldots,N.
	\end{array}
	\right.
\end{equation}
This result reveals that the eigenvalues are independent of the choice of the resetting node $r$. On the other hand, the left eigenvectors of $\mathbf{\Pi}(r;\gamma)$ are \cite{ResetNetworks_PRE2020}

\begin{equation}\label{Eigen_left_reset}
	\left\langle\bar{\psi}_\ell(r;\gamma)\right|=
	\left\{
	\begin{array}{ll}
		\displaystyle
		\left\langle\bar{\phi}_1\right|
		+\sum\limits_{m=2}^N\frac{\gamma}{1-(1-\gamma)\lambda_m}\frac{\left\langle r|\phi_m\right\rangle}{\left\langle r|\phi_1\right\rangle}\left\langle\bar{\phi}_m\right|,\,\,
		 &\ell=1,\\
		\displaystyle
		\left\langle\bar{\phi}_\ell\right| \,\,&  \ell=2,\ldots,N. \\
	\end{array}
    \right.
\end{equation}
Similarly, the right eigenvectors are given by \cite{ResetNetworks_PRE2020}
\begin{equation}\label{Eigen_right_reset}
	\left|\psi_\ell(r;\gamma)\right\rangle=
	\left\{
		\begin{array}{ll}
		\displaystyle
		\left| \phi_1\right\rangle\qquad &\ell= 1,\\
		\displaystyle
		\left|\phi_\ell\right\rangle-\frac{\gamma}{1-(1-\gamma)\lambda_\ell}\frac{\left\langle r|\phi_\ell\right\rangle }{\left\langle r|\phi_1\right\rangle} \left|\phi_1\right\rangle  \,\,&\ell= 2,\ldots,N. \\
	\end{array}
    \right.
\end{equation}
In Eqs. (\ref{Eigen_left_reset})-(\ref{Eigen_right_reset}), $|r\rangle$ denotes the vector with all its components equal to 0 except the $r$-th one, which is equal to 1.
\\[2mm]
With the left and right eigenvectors at hand, one can use the spectral representation
\begin{equation}\label{Pi_prob_powert}
	\mathbf{\Pi}(r;\gamma)=\sum_{l=1}^N\zeta_l(\gamma)\left|\psi_l(r;\gamma)\right\rangle\left\langle\bar{\psi}_l(r;\gamma)\right|.
\end{equation}
In this notation, the occupation probability of the process described by Eq. (\ref{mastermarkov}) is  \cite{ResetNetworks_PRE2020}
\begin{equation}
	P_{ij}(t;r,\gamma)=P_j^\infty(r;\gamma)+\sum_{l=2}^N(1-\gamma)^t\lambda_l^t\left[\left\langle i|\phi_l\right\rangle \left\langle\bar{\phi}_l|j\right\rangle-\gamma\frac{\left\langle r|\phi_l\right\rangle \left\langle\bar{\phi}_l|j\right\rangle}{1-(1-\gamma)\lambda_l} \right], \label{Pijspect}
\end{equation}
where $|i\rangle$ and $|j\rangle$ are defined similarly to $|r\rangle$.
The first term of the r.h.s. in Eq. (\ref{Pijspect}) defines the long time stationary distribution $P_j^\infty(r;\gamma)=\left\langle i\left|\psi_1(r;\gamma)\right\rangle \left\langle\bar{\psi}_1(r;\gamma)\right|j\right\rangle$. Using  Eq. (\ref{Eigen_left_reset}) and $\left|\psi_1(r;\gamma)\right\rangle=\left|\phi_1\right\rangle$ in Eq. (\ref{Eigen_right_reset}), we have  \cite{ResetNetworks_PRE2020}
\begin{equation}\label{Pinfvectors_1}
	P_j^\infty(r;\gamma)=P_i^\infty(0)+\gamma\sum_{l=2}^N\frac{\left\langle r|\phi_l\right\rangle \left\langle\bar{\phi}_l|j\right\rangle}{1-(1-\gamma)\lambda_l}.
\end{equation}
Here we use the identity $P_i^\infty(0)=\left\langle i|\phi_1\right\rangle \left\langle\bar{\phi}_1|j\right\rangle$ for the stationary distribution of the random walk without resetting
\cite{NohRieger2004,MasudaPhysRep2017}.
\\[2mm]
Furthermore, from the occupation probability at finite time $P_{ij}(t;r,\gamma)$, in Eq.~(\ref{Pijspect}), by employing the well-known convolution formula for Markov processes, which relates occupation probabilities to the first-passage time distribution \cite{NohRieger2004,Hughes}, an exact expression for the mean first-passage time $\left\langle T_{ij}(r;\gamma)\right\rangle$ of a random walker starting from node $i$ and reaching node $j$ for the first time, while being subject to stochastic resetting to node $r$, can be derived \cite{ResetNetworks_PRE2020}
\begin{equation}\label{MFPT_resetSM}
	\left\langle T_{ij}(r;\gamma)\right\rangle=\frac{\delta_{ij}}{P_j^\infty(r;\gamma)}+
	\frac{1}{P_j^\infty(r;\gamma)}\sum_{\ell=2}^N\frac{
		\left\langle j|\phi_\ell\right\rangle \left\langle\bar{\phi}_\ell|j\right\rangle-\left\langle i|\phi_\ell\right\rangle \left\langle\bar{\phi}_\ell|j\right\rangle
	}{1-(1-\gamma)\lambda_\ell}.
\end{equation}
In particular, the limit $\gamma\to 0$ recovers the mean first-passage time $\left\langle T_{ij}\right\rangle$ for a random walk without reset \cite{reviewjcn_2021}
\begin{equation}\label{MFPT_normal}
	\left\langle T_{ij}\right\rangle=\frac{1}{P_j^\infty}\left[\delta_{ij}+
	\sum_{\ell=2}^N\frac{
		\left\langle j|\phi_\ell\right\rangle \left\langle\bar{\phi}_\ell|j\right\rangle-\left\langle i|\phi_\ell\right\rangle \left\langle\bar{\phi}_\ell|j\right\rangle
	}{1-\lambda_\ell}\right],
\end{equation}
where $P_j^\infty\equiv P_j^\infty(0)$.
\subsection{Synchronous random walkers on networks}
\label{Sec_General}
In this section, we summarize the notation and main results used in the analytical treatment of simultaneous random walkers on networks without resetting (for further details, we refer the reader to Ref. \cite{Riascos_Multiple2020}). We are interested in the movement of $S$ random walkers on a general connected network with $N$ nodes in the set $\mathcal{V}=\{1,2,\ldots,N\}$. Each walker $s$ follows a transition matrix $\mathbf{W}^{(s)}$ of size $N \times N$, with elements $(\mathbf{W}^{(s)})_{ij} = w^{(s)}_{i\to j}$ defining the probability of hopping from node $i$ to node $j$. At each discrete time step $t = 1, 2, \ldots$, all walkers move independently and simultaneously.
\\[2mm]
A matrix $\hat{\mathcal{W}}$ describing the global activity of these $S$ non-interacting random walkers, under synchronous motion, is given by
\begin{equation}\label{Wmatrix_S}
	\hat{\mathcal{W}} \equiv \bigotimes_{s=1}^S \mathbf{W}^{(s)}
	= \mathbf{W}^{(1)} \otimes \mathbf{W}^{(2)} \otimes \cdots \otimes \mathbf{W}^{(S)},
\end{equation}
where $\otimes$ denotes the tensor product (Kronecker product) of matrices.
\\[2mm]
In addition, it is convenient to introduce the notation $\vec{i} \equiv (i_1, i_2, \ldots, i_S) \in \mathcal{V}^S$, where $i_1, i_2, \ldots, i_S \in {1,2,\ldots, N}$, to describe the positions of each walker, with $i_s$ representing the position (node) of walker $s$ on the network.  Using the canonical basis of $\mathbb{R}^N$, denoted as $\{\left| i\right\rangle\}_{i=1}^N$, the probability $\mathcal{P}(\vec{i},\vec{j};t)$ of finding the $S$ walkers at nodes $\vec{j}$ at time $t$, given that they started from initial positions $\vec{i}$ at $t=0$, is given by
\begin{equation}\label{PMnoninter}
	\mathcal{P}(\vec{i},\vec{j};t)=\prod_{s=1}^S \langle i_s|(\mathbf{W}^{(s)})^t|j_s\rangle=\langle \vec{i}|\hat{\mathcal{W}}^t|\vec{j}\rangle,
\end{equation}
where we use the notation $|\vec{i}\,\rangle\equiv
|i_1,i_2,\ldots,i_S\rangle=|i_1\rangle\otimes|i_2\rangle\otimes \cdots \otimes|i_S\rangle$. The matrix $\hat{\mathcal{W}}$ describes the collective movement of $S$ non-interacting random walkers. Its elements, $\mathcal{W}_{\vec{i} \to \vec{j}}\equiv\langle \vec{i}\,|\hat{\mathcal{W}}|\vec{j}\,\rangle$, define the transition probability from the configuration $\vec{i}$ to a new state $\vec{j}$, corresponding to the positions of all $S$ walkers on the network. 
\\[2mm]
The probability $\mathcal{P}(\vec{i}, \vec{j}; t)$ in Eq. (\ref{PMnoninter}) is formally analogous to the transition probability of a single walker, a correspondence that allows us to analytically address the dynamics of $S$ non-interacting random walkers. This framework is particularly useful for computing the mean number of steps required to reach specific configurations. Throughout, we assume that each transition matrix $\mathbf{W}^{(s)}$ is diagonalizable and that the corresponding walker can reach any node from any initial condition. Under this assumption, the right eigenvectors satisfy $\mathbf{W} ^{(s)}|\phi_i^{(s)}\rangle=\lambda_i^{(s)}|\phi_i^{(s)}\rangle$ for $s=1,2, \ldots, S$, where $\{\lambda_i^{(s)}\}_{i=1}^{N}$ denote the eigenvalues of the transition matrix $\mathbf{W} ^{(s)}$ with the corresponding set of right eigenvectors $\{ |\phi_i^{(s)}\rangle\}_{i=1}^{N}$ \cite{reviewjcn_2021}. In terms of these eigenvectors, it is convenient to define
\begin{equation}
	|\phi_{\vec{i}}\rangle\equiv|\phi_{i_1}^{(1)}\rangle\otimes|\phi_{i_2}^{(2)}\rangle\otimes \cdots \otimes|\phi_{i_S}^{(S)}\rangle=\bigotimes_{s=1}^S |\phi_{i_s}^{(s)}\rangle
\end{equation}
and, combining this definition with Eq.~(\ref{Wmatrix_S})
\begin{equation}\label{SpectM}
	\hat{\mathcal{W}}|\phi_{\vec{i}}\rangle=\zeta_{\vec{i}}|\phi_{\vec{i}}\rangle,
\end{equation}
where, using the definition in Eq.~(\ref{Wmatrix_S}), are obtained the eigenvalues of  $\hat{\mathcal{W}}$
\begin{equation}\label{zeta_S}
	\zeta_{\vec{i}} \equiv \prod_{s=1}^S \lambda_{i_s}^{(s)}.
\end{equation}
We also require the set of left eigenvectors. For the  individual transition matrix $\mathbf{W}^{(s)}$ we have $ \langle \bar{\phi}^{(s)}_i|\mathbf{W}^{(s)}= \lambda_i^{(s)}\langle \bar{\phi}^{(s)}_i|$; then,   $\langle\bar{\phi}_{\vec{i}}| \equiv \bigotimes_{s=1}^S \langle\bar{\phi}_{i_s}^{(s)}|$ satisfies the eigenvector relation
\begin{equation}
	\langle\bar{\phi}_{\vec{i}}|\hat{\mathcal{W}}=
	\zeta_{\vec{i}}\langle\bar{\phi}_{\vec{i}}|.
\end{equation}
In addition, since each set of eigenvectors of $\mathbf{W}^{(s)}$ satisfies $\delta_{ij}=\langle\bar{\phi}^{(s)}_i|\phi^{(s)}_j\rangle$ and $\mathbb{1}=\sum_{l=1}^N |\phi^{(s)}_l\rangle \langle \bar{\phi}^{(s)}_l |$ \cite{reviewjcn_2021}, from the definitions of $|\phi_{\vec{i}}\rangle$ and $\langle\bar{\phi}_{\vec{j}}|$, we have the orthonormalization condition $\langle\bar{\phi}_{\vec{i}}|\phi_{\vec{j}}\rangle=\delta_{i_1,j_1}\delta_{i_2,j_2}\ldots\delta_{i_S,j_S}
\equiv \delta_{\vec{i},\vec{j}}$ and the completeness relation $\sum_{\vec{l}\in \mathcal{V}^S}\left|\phi_{\vec{l}}\right\rangle\left\langle\bar{\phi}_{\vec{l}}\,\right|=\mathbb{1}^{\otimes S}$. 
\\[2mm]
In the following, we denote the largest eigenvalue of $\mathbf{W}^{(s)}$ as $\lambda_1^{(s)}=1$. According to the Perron--Frobenius theorem, this eigenvalue is unique, and its associated eigenvector determines the stationary distribution of each walker via the relation $P^{s,\infty}_{j}=\langle i|\phi_1^{(s)}\rangle \langle\bar{\phi}^{(s)}_1|j\rangle$. Notably, this distribution is independent of the initial node $i$, as $\langle i|\phi_1^{(s)}\rangle$ remains constant \cite{reviewjcn_2021}. We can now express the stationary distribution of $\mathcal{P}(\vec{i} ,\vec{j};t)$ for the $S$-walker system in terms of the eigenvalues and left and right eigenvectors of $\hat{\mathcal{W}}$. From Eq.~(\ref{PMnoninter}) it is obtained \cite{Riascos_Multiple2020}
\begin{equation}
	\mathcal{P}^{\infty}_{\vec{j}}(\vec{i})=\lim_{T\to \infty}\frac{1}{T}\sum_{t=0}^{T} \sum_{\vec{l}\in \mathcal{V}^S}  \zeta_{\vec{l}}^t\langle \vec{i} |\phi_{\vec{l}}\rangle\langle\bar{\phi}_{\vec{l}}|\vec{j}\rangle
	=\sum_{\vec{l}\in \mathcal{V}^S} \delta_{\zeta_{\vec{l}},1}\langle \vec{i} |\phi_{\vec{l}}\rangle\langle\bar{\phi}_{\vec{l}}|\vec{j}\rangle \, .	\label{Pinf_multiple}
\end{equation}
In Eq. (\ref{Pinf_multiple}), it is essential to define the degeneracy of the eigenvalue $\zeta=1$. From Eq. (\ref{zeta_S}), we observe that multiple eigenvectors can be associated with the maximum eigenvalue  $\max_{{\vec{i}}\in \mathcal{V}^S}{\zeta_{\vec{i}}}=1$.  In the following, we define the degeneracy of this eigenvalue as $\kappa\equiv \sum_{\vec{l}\in \mathcal{V}^S} \delta_{\zeta_{\vec{l}},1}$. The value $\kappa=1$ indicates that all initial configurations can transition to any final state within a finite time, meaning the system is irreducible. Conversely, $\kappa>1$ implies the existence of initial conditions that cannot reach certain final states, resulting in a stationary distribution with zero probability for those states. We define the set $\mathcal{D} \equiv \{\vec{l}\in\mathcal{V}^{S}: \zeta_{\vec{l}}=1\}$ and its complement as $\mathcal{D}^\mathrm{c} \equiv \mathcal{V}^S\setminus\mathcal{D}$. The stationary distribution in Eq.~(\ref{Pinf_multiple}) then takes the form
\begin{equation}
	\mathcal{P}^{\infty}_{\vec{j}}(\vec{i})=\sum_{\vec{l}\in \mathcal{D}} \langle \vec{i} |\phi_{\vec{l}}\rangle\langle\bar{\phi}_{\vec{l}}|\vec{j}\rangle .
\end{equation}
In cases with $\kappa>1$, the stationary distribution depends on the initial configuration $\vec{i}$, whereas for $\kappa=1$ we have $\mathcal{P}^{\infty}_{\vec{j}}(\vec{i})=
P^{1,\infty}_{j_1}P^{2,\infty}_{j_2}\cdots P^{S,\infty}_{j_S}$, a probability independent of the initial conditions of the random walkers.
\\[2mm]
With the formalism and notation introduced, it is possible to compute the average time $\langle T(\vec{i};\vec{j})\rangle\equiv\langle T(i_1,i_2,\ldots,i_S;j_1,j_2,\ldots,j_S)\rangle $ required for the walkers to simultaneously reach, for the first time, the nodes described by the vector $\vec{j}$, given that they start at nodes $\vec{i}$ at time $t=0$. A detailed derivation of this result can be found in Ref. \cite{Riascos_Multiple2020}, leading to the following expression for $\vec{i}\neq \vec{j}$
\begin{equation}\label{TijSpect}
	\langle T(\vec{i};\vec{j}\,)\rangle
	=\frac{1}{\mathcal{P}_{\vec{j}}^\infty(\vec{i}\,)}
	\sum_{\vec{l}\in \mathcal{D}^\mathrm{c}}\frac{\langle \vec{j} |\phi_{\vec{l}}\rangle\langle\bar{\phi}_{\vec{l}}|\vec{j}\rangle-
		\langle \vec{i} |\phi_{\vec{l}}\rangle\langle\bar{\phi}_{\vec{l}}|\vec{j}\rangle}{1-\zeta_{\vec{l}}}
\end{equation}
and $\langle T(\vec{i};\vec{i}\,)\rangle=1/\mathcal{P}_{\vec{i}}^\infty(\vec{i}\,)$.
\subsection{Mean first-encounter times for two normal random walkers}
The result in Eq.~(\ref{TijSpect}) allows the analysis of the mean first-encounter times for $S$ non-interacting random walkers on networks by considering $\vec{j}=(j,j,\ldots,j)$. In this section, we apply 
Eqs.~(\ref{Pinf_multiple})--(\ref{TijSpect}) to compute the mean time $\langle T(i_1,i_2;j,j)\rangle$ (we refer to this quantity as mean first-encounter time) required for two ($S=2$) random walkers, initially positioned at nodes $i_1$ and $i_2$ at $t=0$, to meet for the first time at node $j_1=j_2=j$. Each walker moves according to an individual transition probability matrix $\mathbf{W}$, defined in terms of the adjacency matrix elements $A_{lm}$ as $w_{l\to m} \equiv A_{lm}/k_l$, where $k_l \equiv \sum_{m=1}^N A_{lm}$ denotes the degree of node $l$ (we refer to this dynamics as {\it normal random walk}). The stationary distribution of a single walker is given by $P_{j}^{\infty}=\frac{k_j}{\sum_{l=1}^N k_l}$ \cite{NohRieger2004}. 
\\[2mm]
Furthermore, the degeneracy of the largest eigenvalue is given by $\kappa=\sum_{l,m=1}^N \delta_{\lambda_l\lambda_m,1}$. Consequently, we obtain $\kappa=2$ if the transition matrix has eigenvalues $\lambda=\pm 1$. For normal random walks, this occurs in bipartite networks \cite{VanMieghem2011,FractionalBook2019}, a particular class of undirected graphs where the vertex set can be partitioned into two disjoint subsets, such that each edge connects nodes belonging to different subsets. Examples include cycles with an even number of nodes and trees. If the network is not bipartite, then $\kappa=1$ (corresponding to $\lambda=1$), and the stationary distribution in Eq. (\ref{Pinf_multiple}) simplifies to $\mathcal{P}^{\infty}_{\vec{j}}(\vec{i}\,)=(P_{j}^{\infty})^2$, independent of the initial node. Therefore, 
Eq.~(\ref{TijSpect}) gives \cite{Riascos_Multiple2020}
\begin{equation}\label{MFETtworandomkappa1}
	\fl
	\qquad
	\langle T(i_1,i_2;j,j)\rangle =\frac{1}{(P_{j}^\infty)^2}
	\Big[\delta_{i_1j}\delta_{i_2j}+\sum_{l,m=1}^N g(\lambda_l\lambda_m)\left(X_{jj}^{(l)}X_{jj}^{(m)}-X_{i_1j}^{(l)}X_{i_2j}^{(m)}\right)\Big]
\end{equation}
where we use the notation
\begin{equation}
	X_{ij}^{(l)}\equiv \left\langle i|\phi_l\right\rangle \left\langle \bar{\phi}_l|j\right\rangle
\end{equation} 
and 
\begin{equation}
	g(z) \equiv \left\{
	\begin{array}{ll}
		\displaystyle
		(1 - z)^{-1} & \qquad \mathrm{for}\ z \neq 1, \\
		0            & \qquad \mathrm{for}\ z = 1.
	\end{array}
	\right.
\end{equation}
The formalism presented in this section offers an analytical approach to characterize synchronous motion through the spectral decomposition of diagonalizable transition matrices, assuming independent Markovian random walkers.
\section{Simultaneous random walkers and the effect of reset}
\label{Sec_Simul_RW_reset}
In the previous sections, we presented results on the resetting dynamics of random walkers on networks \cite{ResetNetworks_PRE2020}, as well as the formalism for studying the encounter times of multiple simultaneous random walkers on networks \cite{Riascos_Multiple2020}. Building upon this theoretical framework, we now establish the foundation for deriving expressions that characterize the encounter times of multiple walkers in a network, where one or more follow a resetting strategy. In this section, we derive analytical expressions to investigate the effect of resetting on the mean first-encounter times of two random walkers. Furthermore, we explore various examples to examine the impact of resetting for different network topologies, including rings, Cayley trees, and random networks generated using the  Erdős-Rényi, Watts-Strogatz and, Barabási-Albert algorithms.

\subsection{Two random walkers: one with resetting and one normal}
In order to derive an expression for the mean first-encounter times in the case of one normal random walker with resetting and the second one following the normal random walk strategy, we start from Eq. (\ref{MFETtworandomkappa1}). The effect of resetting must be incorporated in the first walker, replacing the factors $X_{ij}^{(l)}$ with new terms $Y_{ij,r}^{(l)}(\gamma)$, which explicitly account for the resetting effects.  Additionally, the eigenvalues and stationary distribution are substituted by their counterparts in the resetting walk, as given by Eqs. (\ref{eigvals_zeta}) and (\ref{Pinfvectors_1}), respectively. Consequently, the mean first-encounter time at node $j$ is given by

\begin{eqnarray}\label{Tiij_XY}
	\fl
	\langle T(i_1,i_2,r,j) \rangle = \frac{1}{P_j^\infty(r;\gamma) P_j^\infty } \left[ \delta_{i_1j} \delta_{i_2j} + \sum_{l,m=1}^N g(\zeta_l(\gamma) \lambda_m ) \left(Y_{jj,r}^{(l)}X_{jj}^{(m)} - Y_{i_1j,r}^{(l)}X_{i_2j}^{m)} \right) \right],
\end{eqnarray}
with $X_{ij}^{(m)}\equiv \langle i | \phi_m \rangle \langle \bar{\phi}_m | j \rangle$ and $Y_{ij,r}^{(l)}(\gamma) \equiv \langle i | \psi_l(r;\gamma) \rangle \langle \bar{\psi}_l(r;\gamma) | j \rangle$. The values $Y_{ij,r}^{(l)}(\gamma)$ are expressed in terms of the eigenvectors with resetting, given by Eqs. (\ref{Eigen_left_reset}) and (\ref{Eigen_right_reset}). Here, $r$ denotes the reset node of the first random walker, and $\gamma$ represents the probability that the walker resets to node $r$. Therefore, using Eqs. (\ref{Eigen_left_reset}) and (\ref{Eigen_right_reset}), $Y_{ij,r}^{(l)}(\gamma)$ can be expressed in terms of $X_{ij}^{(l)}$ as
\begin{equation}\label{Yijm}
	Y_{ij,r}^{(l)}(\gamma)=\left\{
	\begin{array}{ll}
		\displaystyle
		X_{ij}^{(1)}+ \gamma \sum^N_{k=2} \frac{X^{(k)}_{rj}}{1-(1-\gamma)\lambda_k}&l= 1,\\
	\displaystyle	X_{ij}^{(l)} - \gamma  \frac{X^{(l)}_{rj}}{1-(1-\gamma)\lambda_l} &l= 2,\ldots,N.\\
	\end{array}
	\right.
\end{equation}
Then, Eq. (\ref{Yijm}) can be incorporated into Eq. (\ref{Tiij_XY}) (see Appendix \ref{Ref_Sec_Append_1} for a detailed derivation), leading to the expression for the mean first-encounter time
\begin{eqnarray}\nonumber
	\fl
	\langle T(i_1,i_2,r,j) \rangle =& \frac{1}{P_j^\infty(r;\gamma)P_j^\infty } \left[  \delta_{i_1j} \delta_{i_2j} + \sum_{l,m=1}^N g(\zeta_l(\gamma) \lambda_m) \left(X_{jj}^{(l)}X_{jj}^{(m)} - X_{i_1j}^{(l)}X_{i_2j}^{(m)} \right) \right. \\ 
	& \left. + \gamma \sum_{l,m=2}^N \frac{\lambda_m}{(1-\lambda_m)(1-(1-\gamma)\lambda_l\lambda_m)}\left(X_{rj}^{(l)}X_{jj}^{(m)} - X_{rj}^{(l)}X_{i_2j}^{(m)} \right)  \right].\label{T_encounter_reset1_normal2}
\end{eqnarray}
This analytical expression provides the mean first-encounter time for two walkers: the first one follows a resetting strategy to node $r$, while the second one follows a normal random walk strategy. The results are expressed in terms of the known properties of the normal random walk. The stationary distribution $P_j^\infty(r;\gamma)$ for the random walker with reset is given by Eq. (\ref{Pinfvectors_1}), while $\zeta_l(\gamma)$ is given by Eq. (\ref{eigvals_zeta}). Furthermore, the result can be expressed in terms of $\lambda_k$, the eigenvalues of the normal random walk. In addition, in Eq. (\ref{T_encounter_reset1_normal2}), it can be observed that setting $\gamma = 0$ recovers the expression in Eq. (\ref{MFETtworandomkappa1}) for two walkers following a normal random walk strategy, since $\zeta_l(0) = \lambda_l$.
\subsubsection{Encounter times on rings:}
\begin{figure}[!t]
	\centering
	\includegraphics[width=1.0\textwidth]{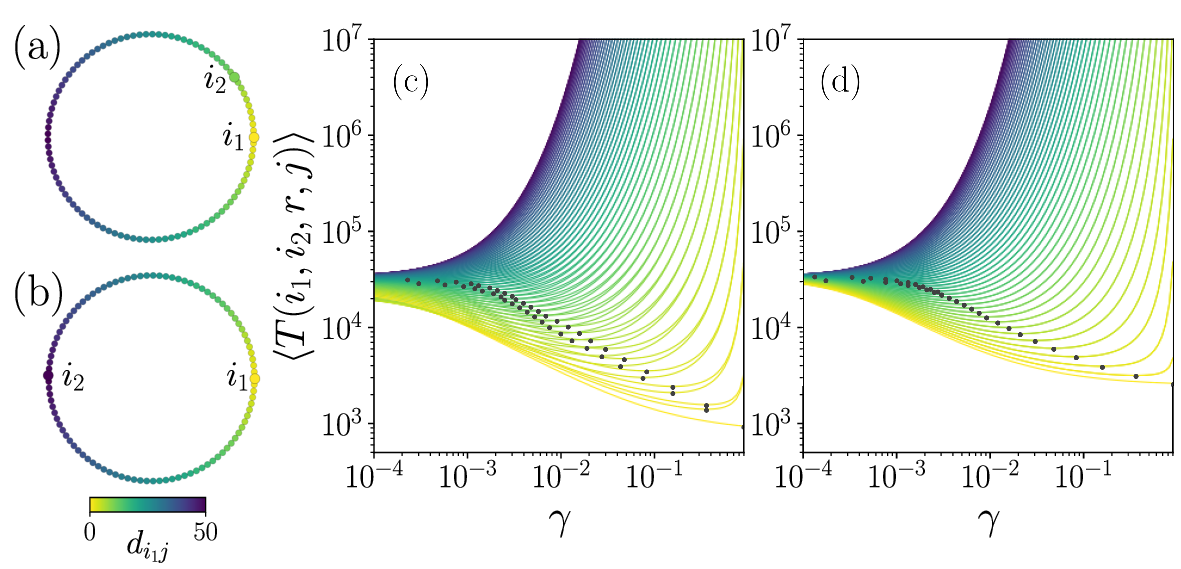}
	\caption{Mean first-encounter times at node $j$ for two walkers as a function of the reset probability $\gamma$ on a ring with $N=101$ nodes. The first walker follows a movement with resets to its initial node $r=i_1$, while the second follows a normal random walk strategy starting from node $i_2$. Two cases are analyzed: (a) with $d_{i_1\, i_2}=10$ and (b) using $d_{i_1\,i_2}=50$. In (c) and (d), we present the respective curves of $\langle T(i_1,i_2,r,j) \rangle$ for different nodes $j$ where the random walkers meet for the first time, as a function of $\gamma$, corresponding to the cases in (a) and (b). The colors of the curves represent the distance $d_{i_1 j}$ and are encoded in the color bar. The dots in the curves represent the values where  $\langle T(i_1,i_2,r,j) \rangle$ is minimum.  }
	\label{Fig_2}
\end{figure}

Let us now analyze the effects of resetting on the first random walker while maintaining the normal random walk strategy for the second one, considering the simultaneous dynamics on a ring with $N$ nodes. In this network, $k_i = 2$, and the transition matrix without resetting $\mathbf{W}$ is a circulant matrix with well-known eigenvalues and eigenvectors \cite{VanMieghem2011, FractionalBook2019}. We limit the discussion to the case of rings with an odd number of nodes $N$; in this case the network is not bipartite and $\kappa=1$ for the dynamics without reset $\gamma=0$, allowing the application of Eq. (\ref{T_encounter_reset1_normal2}) for $0\leq \gamma <1$.
\\[2mm]
The eigenvalues of $\mathbf{W}$ are given by $\lambda_{l}=\cos\varphi_{l}$, with $\varphi_{l} \equiv \frac{2\pi}{N}(l-1)$. The projections of the eigenvectors in the canonical basis are $\langle j|\phi_{l}\rangle=\frac{1}{\sqrt{N}}e^{-\mathbf{i}\varphi_{l}(j-1)}$ and $\langle \bar{\phi}_{l}|j\rangle=\frac{1}{\sqrt{N}}e^{\mathbf{i}\varphi_{l}(j-1)}$ \cite{VanMieghem2011}, where $\mathbf{i}=\sqrt{-1}$.
Therefore, incorporating this information into Eq. (\ref{T_encounter_reset1_normal2}), we have
\begin{eqnarray}\nonumber
& \fl	\langle T(i_1,i_2,r,j) \rangle = \frac{1}{P_j^\infty(r;\gamma) P_j^\infty } \left[  \delta_{i_1 j} \delta_{i_2 j} + \sum_{l,m=1}^N g(\zeta_l(\gamma) \lambda_m ) \left(\frac{1-e^{\mathbf{i}[\varphi_l(j-i_1)+\varphi_m(j-i_2)] }}{N^2} \right) \right. \\  
	& \fl	\qquad \left. + \gamma \sum_{l,m=2}^N \frac{\cos \varphi_m}{\left(1-\cos\varphi_m\right)\left(1-(1-\gamma)\cos \varphi_l \cos\varphi_m\right)}\left(\frac{1-e^{\mathbf{i}\varphi_m(j-i_2)}}{N^2} \right)e^{\mathbf{i}\varphi_l(j-r)}   \right]. \label{Tiijr_1reinicio_anillo}
\end{eqnarray}
From Eq. (\ref{Pinfvectors_1}), the stationary distribution $P_j^\infty(r;\gamma)$ for the dynamics with reset on rings takes the form
\begin{equation}
P_j^\infty(r;\gamma)=\frac{1}{N}+\frac{\gamma}{N}\sum_{m=2}^N  \frac{\cos(\varphi_m d_{jr})}{1-(1-\gamma)\cos\varphi_m}
\end{equation}
where $d_{ij}$ is the distance between the nodes  $i$ and $j$, whereas $P_j^\infty=\frac{1}{N}$ for the normal random walk.
\\[2mm]
In Fig. \ref{Fig_2}, we present the numerical results for $\langle T(i_1,i_2,r,j) \rangle$ as a function of $\gamma$. These values are obtained using Eq.~(\ref{Tiijr_1reinicio_anillo}) for the simultaneous dynamics on a ring of size $N=101$, considering two initial node configurations: one in Fig. \ref{Fig_2}(a) with $d_{i_1 i_2} = 10$ and another in Fig. \ref{Fig_2}(b) with $d_{i_1 i_2} = 50$. The different curves correspond to all possible nodes $j=1,2,\dots,N$ where the random walkers meet for the first time, the color bar codifies the distance $d_{i_1 j}$. Figures \ref{Fig_2}(c)–(d) show the respective results for the configurations in Figs. \ref{Fig_2}(a)–(b). In these cases, the first walker resets stochastically to its initial node $r=i_1$, while the second walker, following a normal random walk strategy, starts at node $i_2$.
\\[2mm]
The results obtained for the case with $d_{i_1 i_2}=10$ in Fig. \ref{Fig_2}(a), with mean first-encounter times $\langle T(i_1,i_2,r,j) \rangle$ reported in Fig. \ref{Fig_2}(c), show that in the limit $\gamma\to 0$, the encounter times take average values in the interval $18661<\langle T(i_1,i_2,r,j) \rangle<35256$. However, when reset is included with values $\gamma\in [10^{-4},1)$, the results exhibit drastic modifications depending on the node $j$. In particular, for $d_{i_1 j}\leq 23$, the reset reduces the mean first-encounter times [the minimum values of $\langle T(i_1,i_2,r,j) \rangle$ found for the optimal reset are presented with a dot in Fig. \ref{Fig_2}(c)]. In contrast, for nodes at distances $d_{i_1 j}>23$, the reset always increases $\langle T(i_1,i_2,r,j) \rangle$ compared to the times found for the dynamics without reset, showing that in this case, the reset hinders the coincidence of the walkers at these specific nodes. A similar behavior is observed for the initial nodes illustrated in Fig. \ref{Fig_2}(b), with mean first-encounter times $\langle T(i_1,i_2,r,j) \rangle$ presented in Fig. \ref{Fig_2}(d). In this case, the effect of resetting the first random walker also reduces the mean first-encounter times for target nodes closer to the reset node. However, for $j$ with $d_{i_1 j} > 25$, the reset always increases the mean first-encounter times.
\\[2mm]
In this way, the observed results are intuitive. Without resetting, the two walkers may take a long time to coincide, as they are free to move across the entire ring. By introducing resetting in the first walker, it remains longer in nodes near the initial node $i_1$, increasing the likelihood that the second walker encounters it more quickly at nodes $j$ close to $i_1$, while significantly delaying encounters at nodes farther from the resetting node.
\subsubsection{Encounter times on different structures:}
\begin{figure*}[!t]
	\centering
	\includegraphics[width=1.0\textwidth]{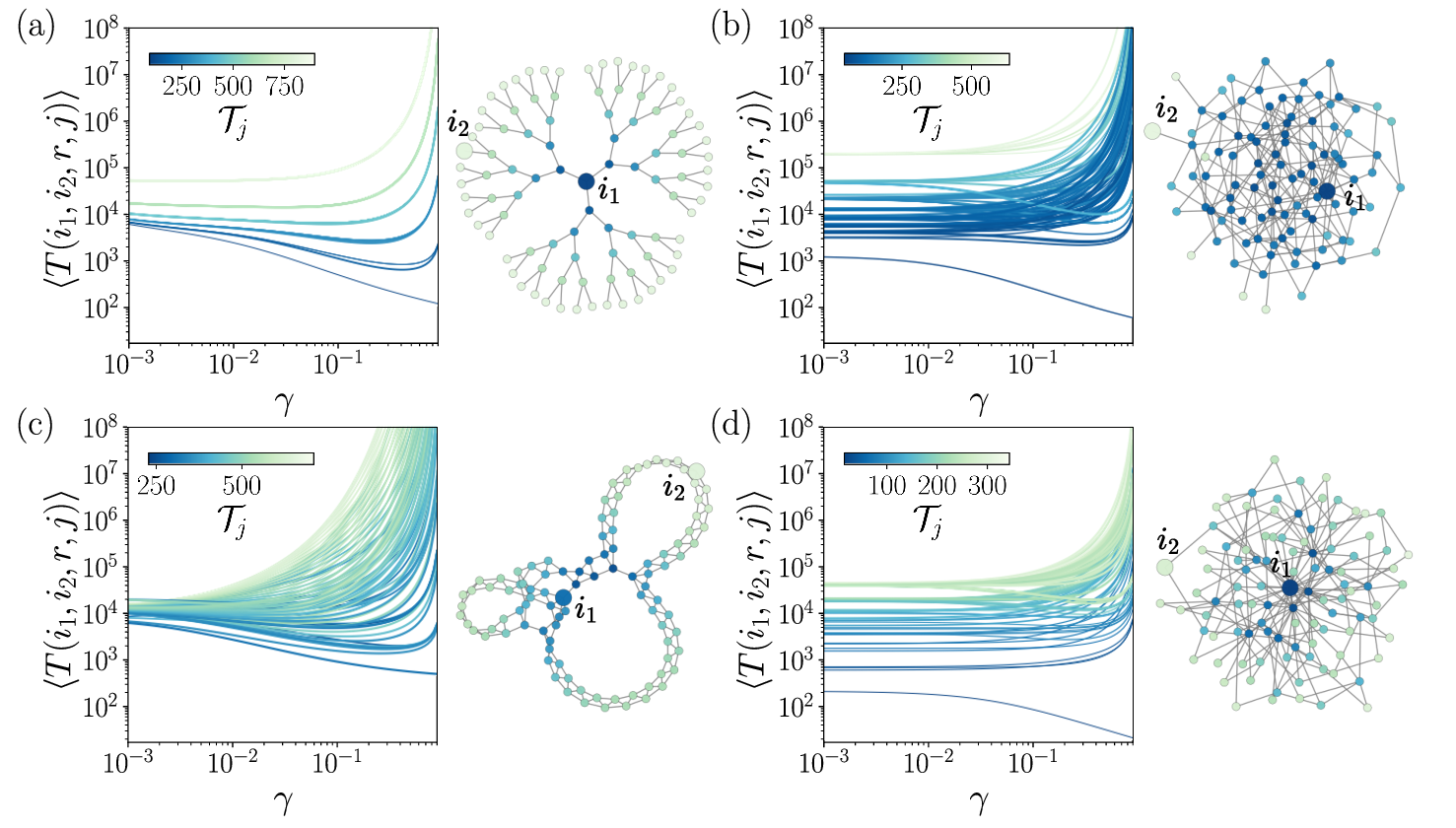}
	\caption{Mean first-encounter times $ \langle T(i_1,i_2,r,j) \rangle $ for simultaneous random walks on networks: (a) Cayley tree; (b) Erdős-Rényi; (c) Watts-Strogatz; and (d) Barab\'asi-Albert random networks. The first walker follows a movement that restarts stochastically with probability $ \gamma $ to its initial node $ r=i_1 $, while the second follows a normal random walk strategy starting from node $ i_2 $ (the right panels show the network and the initial nodes $i_1$, $i_2$ analyzed). We depict the global time $ \langle T(i_1,i_2,r,j) \rangle $ as a function of $ \gamma $ for all nodes $ j=1,\ldots,N $, where the random walkers coincide for the first time. To identify the effects of resetting, each node $ j $ and its corresponding curve are colored according to the global time $ \mathcal{T}_j\equiv \frac{1}{N}\sum_{i=1}^N\left\langle T_{ij}\right\rangle $, where $\left\langle T_{ij}\right\rangle$ is defined in Eq. (\ref{MFPT_normal}).}
	\label{Fig_3}
\end{figure*}
With the help of Eq. (\ref{T_encounter_reset1_normal2}), we further analyze the mean first-encounter times $ \langle T(i_1,i_2,r,j) \rangle $ for simultaneous random walks on different types of networks of relatively small size for clarity in the visualizations. The first walker resets with probability $ \gamma $ to its initial node $ r=i_1 $, while the second follows a normal random walk strategy starting from node $ i_2 $.
\\[2mm]
In Fig. \ref{Fig_3}(a), we analyze encounter times in a finite Cayley tree with coordination number $ z=3 $ and composed of $ n=5 $ shells. The nodes of the last shell have degree $ 1 $, whereas the other nodes have degree $ z $. Similarly, the studied network in Fig. \ref{Fig_3}(b) is an Erdős-Rényi random network   \cite{ErdosRenyi1959} with Poisson degree distribution and average degree $\langle k \rangle= 2.72$. Figure~\ref{Fig_3}(c) shows $ \langle T(i_1,i_2,r,j) \rangle $ for a Watts-Strogatz network \cite{WattsStrogatz1998} generated from a ring with nearest-neighbor and next-nearest-neighbor links and a rewiring probability of $ p=0.01 $. The shortcuts break the translational invariance of the ring geometry. Figure \ref{Fig_3}(d) corresponds to the analysis of a scale-free Barabási-Albert network generated with the preferential attachment rule with $ m=2 $ \cite{BarabasiAlbert1999}.
\\[2mm]
The results obtained for the different networks analyzed in Fig.~\ref{Fig_3} reveal distinct behaviors when resetting is introduced, depending on the specific node $j$ where the encounters occur. Nevertheless, it is important to emphasize that, across all network types, there are multiple cases in which resetting the first random walker to its initial node effectively reduces the mean first-encounter times. This indicates that stochastic resetting can facilitate encounters by increasing the likelihood of both walkers meeting at particular target nodes.
\subsection{Two random walkers with resetting to the initial node}
Similarly to the case where one walker follows a resetting strategy while the other performs a normal random walk, the resetting dynamics can be incorporated into both walkers by using the factors $Y_{ij,r}^{(m)}(\gamma)$, along with the eigenvectors and eigenvalues of the resetting process given in 
Eqs.~(\ref{eigvals_zeta})--(\ref{Eigen_right_reset}). This approach allows us to consider both walkers experiencing resetting with probabilities $\gamma_1$ and $\gamma_2$ to nodes $r_1$ and $r_2$, respectively. The mean first-encounter time is given by
\begin{eqnarray}
	\nonumber
	\displaystyle
	\langle T(\vec{i},\vec{r},j) \rangle = \frac{1}{P_{j}^\infty(r_1;\gamma_1) P_{j}^\infty(r_2;\gamma_2)} \Bigg[ \delta_{i_1j} \delta_{i_2j}\\ \hspace{3cm} + \sum_{l,m=1}^N g(\zeta_l(\gamma_1) \zeta_m(\gamma_2)) \left(Y_{jj,r_1}^{(l)}Y_{jj,r_2}^{(m)} - Y_{i_1j,r_1}^{(l)}Y_{i_2j,r_2}^{(m)} \right) \Bigg], \label{Tiij_YY}
\end{eqnarray}
where the values of $Y_{ij,r}^{(m)}(\gamma)$ are given by Eq. (\ref{Yijm}).
Substituting these expressions into Eq. (\ref{Tiij_YY}), we obtain the following result (see Appendix \ref{Ref_Sec_Append_2} for details)
\begin{eqnarray}\nonumber
\fl	\langle T(\vec{i},\vec{r},j) \rangle = \frac{1}{P_{j}^\infty(r_1;\gamma_1) P_{j}^\infty(r_2;\gamma_2) } \Bigg[ \delta_{i_1j} \delta_{i_2j} + \sum_{l,m=1}^N  g(\zeta_l(\gamma_1) \zeta_m(\gamma_2) ) \left(X_{jj}^{(l)}X_{jj}^{(m)} - X_{i_1j}^{(l)}X_{i_2j}^{(m)} \right)   \\  \nonumber \fl	\qquad +  \gamma_1 \sum_{l,m=2}^N  \frac{\lambda_m(1-\gamma_2)}{(1-(1-\gamma_2)\lambda_m)(1-(1-\gamma_1)(1-\gamma_2)\lambda_l\lambda_m)}\left(X^{(l)}_{r_1j}X_{jj}^{(m)}-X^{(l)}_{r_1j}X_{i_2j}^{(m)} \right)  \\  \fl	\qquad +  \gamma_2 \sum_{l,m=2}^N \frac{\lambda_l(1-\gamma_1)}{(1-(1-\gamma_1)\lambda_l)(1-(1-\gamma_1)(1-\gamma_2)\lambda_l\lambda_m)}\left(X^{(l)}_{jj}X_{r_2j}^{(m)}-X^{(l)}_{i_1j}X_{r_2j}^{(m)} \right)\Bigg]. \label{Tiijr_2reinicio}
\end{eqnarray}
In this equation, it can be observed that setting the resetting parameter $\gamma_2 = 0$ recovers the expression for the case where only one walker undergoes resetting. Furthermore, when $\gamma_1 = \gamma_2 = 0$, the result reduces to the expression for two normal walkers given in Eq. (\ref{MFETtworandomkappa1}). Additionally, this formulation expresses the mean first-encounter time in terms of the information of a normal random walk strategy, implying that, given the eigenvalues and eigenvectors of a normal walker, one can determine the mean first-encounter times for the dynamics with reset.
\begin{figure*}[t]
	\centering
	\includegraphics[width=1.0\textwidth]{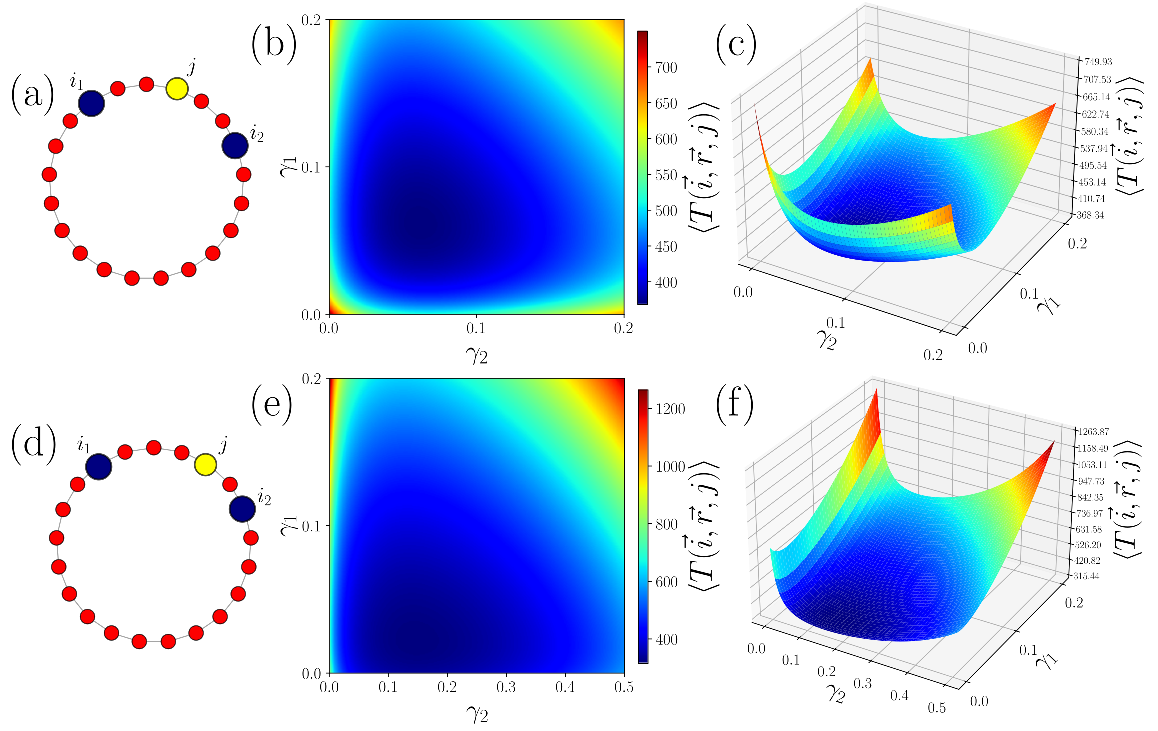}
	\vspace{-6mm}
	\caption{Mean first-encounter times of two walkers with resetting to their initial nodes on a ring with $N = 21$ nodes. (a) Schematic representation of the ring graph, illustrating the initial positions $i_1$ and $i_2$ of the two walkers, which reset to these nodes with probabilities $\gamma_1$ and $\gamma_2$, respectively. The target node $j$ is located symmetrically between $i_1$ and $i_2$. (b) Heatmap of the mean first-encounter times $\langle T(\vec{i},\vec{r},j) \rangle$ at node $j$ as a function of the resetting probabilities $\gamma_1$ and $\gamma_2$, with the color bar encoding the corresponding values. (c) Surface plot showing the same data as in panel (b), illustrating the behavior of $\langle T(\vec{i},\vec{r},j) \rangle$ across the full range of resetting probabilities. (d)–(f) Analogous analysis performed for an asymmetric configuration where the target node $j$ is positioned closer to walker $i_1$, breaking the symmetry with respect to the initial conditions.}
	\label{Fig_4}
\end{figure*}
\begin{figure*}[t!]
	\centering
	\includegraphics*[width=1.0\textwidth]{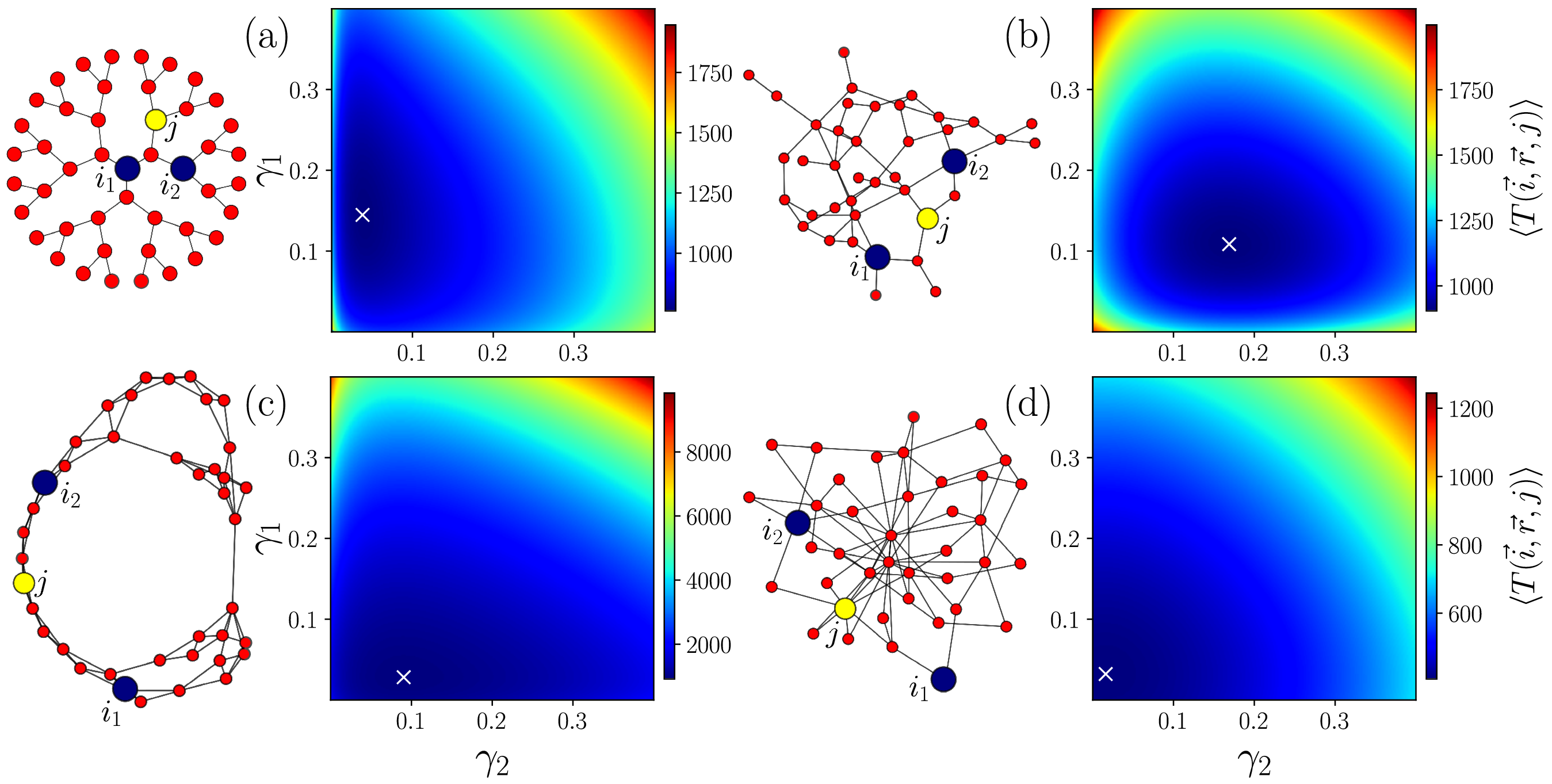}
	\caption{Mean first-encounter times of two walkers with resetting to their initial nodes on networks.
		(a) Cayley tree with coordination number $z = 3$ and depth $n = 4$. The network illustrates the initial positions $i_1$ and $i_2$  of the two walkers, each resetting to their respective initial nodes with probabilities $\gamma_1$ and $\gamma_2$. The target node $j$ is also indicated. The heatmap shows the mean first-encounter times $\langle T(\vec{i},\vec{r},j) \rangle$, encoded in the color bar, at node $j$ as a function of the resetting probabilities. The marker ``$\times$'' shows the values for which the encounter time is minimized.
		Panels (b)–(d) present the same analysis for random networks generated using the following models: (b) Erdős–Rényi, (c) Watts–Strogatz, and (d) Barabási–Albert.}
	\label{Fig_5}
\end{figure*}
\\[2mm]
To illustrate the results of Eq. (\ref{Tiijr_2reinicio}), in Figs. \ref{Fig_4}–\ref{Fig_5} we present an analysis of the mean first-encounter times $\langle T(\vec{i},\vec{r},j) \rangle$ for two walkers with resetting, where the resetting nodes coincide with the initial positions of the walkers, i.e., $r_s = i_s$ for $s = 1, 2$. Figure~\ref{Fig_4} shows two scenarios involving walkers on a ring with $N = 21$ nodes. In the first case, illustrated in Figs. \ref{Fig_4}(a)–(c), the configuration consists of two walkers starting at nodes $i_1$ and $i_2$, with the target node $j$ placed symmetrically between the two starting positions, such that $d_{i_1j} = d_{i_2j} = 3$, as depicted in Fig. \ref{Fig_4}(a). Figures \ref{Fig_4}(b)–(c) show the mean first-encounter times as a function of the resetting probabilities $\gamma_1$ and $\gamma_2$. Both visualizations reveal a clear minimum at $\gamma_1 = \gamma_2 = 0.0613$ where $\langle T(\vec{i},\vec{r},j) \rangle=368.34$, highlighting how resetting dynamics can significantly reduce the mean first-encounter times in comparison to the times for the dynamics without resetting, where the time takes the value $\langle T(\vec{i},\vec{r},j) \rangle = 654.02$. In the second case [Figs.~\ref{Fig_4}(d)–(e)], the effect of resetting is examined in a less symmetric configuration, where the target node $j$ is positioned closer to node $i_1$. In this setup, the distances are $d_{i_1j} = 4$ and $d_{i_2j} = 2$. As a result of this asymmetry, the symmetry observed in the previous case is broken, and a new minimum in the mean first-encounter time is obtained at $\gamma_1 = 0.022$ and $\gamma_2 =0.146$.
\\[2mm]
Furthermore, based on the analytical result in Eq.~(\ref{Tiijr_2reinicio}), Fig.~\ref{Fig_5} presents a study of the encounter dynamics with resetting for different network topologies. We show the analyzed network along with the initial positions $i_1$ and $i_2$ of the two walkers, each resetting to their respective initial nodes with probabilities $0<\gamma_1,\gamma_2<0.4$, as well as the target node $j$. The heatmaps display $\langle T(\vec{i}, \vec{r}, j) \rangle$ as a function of the resetting probabilities.
\\[2mm]
Figure~\ref{Fig_5}(a) depicts a Cayley tree with coordination number $z=3$ and depth $n=4$. In this configuration, one walker resets to the central node $i_1$ of the tree, while the second walker resets to node $i_2$, located at the second layer. The target node $j$ is also positioned in the second layer, within the same branch as $i_2$. The results show that resetting the first walker to the central node significantly facilitates the encounter, and resetting the second walker also contributes, though to a lesser extent. The mean first-encounter times, shown for this case, exhibit a clear minimum as a function of the resetting parameters $\gamma_1 > 0$ and $\gamma_2 > 0$ (the minimum is marked with an ``$\times$’’).
\\[2mm]
In Figs.~\ref{Fig_5}(b)--(d) we repeat the same analysis for complex networks with $N=40$ generated randomly with the algorithms of Erdős–Rényi with a probability to establish a link $p=0.078$ [in Fig.~\ref{Fig_5}(b)], Watts–Strogatz with rewiring probability $p=0.1$ [in Fig.~\ref{Fig_5}(c)], and Barabási–Albert obtained by adding $m=2$ at each step of the preferential attachment algorithm [in Fig.~\ref{Fig_5}(d)]. All the results show cases where non-zero values of the resetting probabilities $\gamma_1$ and $\gamma_2$ minimize the mean first-encounter times of two walkers, $\langle T(\vec{i}, \vec{r}, j) \rangle$. 
\\[2mm]
Our findings in Figs.~\ref{Fig_4} and \ref{Fig_5} reveal specific scenarios in which stochastic resetting serves as an effective mechanism for reducing the mean first-encounter times between two random walkers on a network. For both regular structures, such as rings and Cayley trees, and heterogeneous random networks, resetting to the initial positions of the walkers results in a clear optimization of the encounter dynamics. Notably, the presence of a well-defined minimum in $\langle T(\vec{i}, \vec{r}, j) \rangle$ for intermediate values of $\gamma_1$ and $\gamma_2$ suggests that neither vanishing nor excessively frequent resetting is optimal. These findings emphasize the relevance of tuning resetting strategies to improve search and encounter processes in networks with diverse topologies. 
\section{Conclusions}
\label{Sec_Conclusions}
In this research, we investigate the dynamics of simultaneous random walkers with resetting on networks. We derive two exact analytical expressions for the mean first-encounter times of two Markovian walkers. The first expression corresponds to the case where only one walker employs a resetting strategy, while the second addresses the scenario in which both walkers reset to their initial positions. In both cases, 
the mean encounter-times are expressed in terms of the eigenvalues and eigenvectors of the transition matrix associated with the normal random walk. This formulation allows the impact of resetting on encounter times to be understood directly from the spectral properties of the underlying dynamics without resetting. We illustrate our findings with different cases where the resetting dynamics significantly reduce the mean first-encounter times on rings, Cayley trees, and random networks generated using the Erdős--Rényi, Watts--Strogatz, and Barabási--Albert algorithms. Our findings reveal specific cases in which stochastic resetting serves as an effective mechanism for reducing the mean first-encounter times between two random walkers on a network. For both regular structures, such as rings and Cayley trees, and heterogeneous random networks, resetting to the initial positions of the walkers may optimize the encounter dynamics. These results also highlight the importance of properly tuning the resetting strategies to improve encounter times in networks with diverse topologies.
\\[2mm]
The formalism introduced in this work can be extended to the study of simultaneous processes with resetting involving other types of random walkers and diffusive transport dynamics on networks. It also opens the possibility for analyzing scenarios with more than two simultaneous random walkers. The analytical framework and results presented may prove useful in diverse contexts, including human mobility, epidemic spreading, foraging behavior, and other applications where encounter dynamics play a central role.
\section{Appendices}
\subsection{Derivation of Eq.~(\ref{T_encounter_reset1_normal2})}
\label{Ref_Sec_Append_1}
In this appendix, we derive Eq. (\ref{T_encounter_reset1_normal2}) starting from Eq. (\ref{Tiij_XY}). Our objective is to obtain an analytical expression for the mean first-encounter time of two random walkers, where the first walker follows a resetting strategy, while the second performs a normal (non-resetting) random walk. We begin by considering Eq.~(\ref{Tiij_XY}) as the starting point of our derivation
\begin{equation}\label{Tiij_XY_apendix}
	\fl
	\langle T(i_1,i_2,r,j) \rangle = \frac{1}{P_j^\infty(r;\gamma) P_j^\infty } \left[ \delta_{i_1j} \delta_{i_2j} + \sum_{l,m=1}^N g(\zeta_m(\gamma) \lambda_l ) \left(Y_{jj,r}^{(l)}X_{jj}^{(m)} - Y_{i_1j,r}^{(l)}X_{i_2j}^{(m)} \right) \right],
\end{equation}
where $Y_{ij,r}^{(l)}$ is given by
\begin{equation}\label{Yijm_appendix_1}
	Y_{ij,r}^{(l)}(\gamma)=
	\left\{
	\begin{array}{ll}
		\displaystyle
		X_{ij}^{(1)}+ \gamma \sum^N_{k=2} \frac{X^{(k)}_{rj}}{1-(1-\gamma)\lambda_k}&l= 1,\\
		X_{ij}^{(l)}- \gamma  \frac{X^{(l)}_{rj}}{1-(1-\gamma)\lambda_l} &l= 2,\ldots,N. \\
	\end{array}
    \right.
\end{equation}
We substitute this expression into Eq.~(\ref{Tiij_XY_apendix}). To proceed, we distinguish between two cases: when $l = 1$, and when $l = 2, \ldots, N$. Accordingly, we split the summation into these two contributions as follows
\begin{eqnarray}\nonumber
	\fl
	\langle T(i_1,i_2,r,j) \rangle = \frac{1}{P_j^\infty(r;\gamma) P_j^\infty } \left[ \delta_{i_1j} \delta_{i_2j} + \sum_{m=1}^N g(\zeta_1(\gamma) \lambda_m ) \left(Y_{jj,r}^{(1)}X_{jj}^{(m)} - Y_{i_1j,r}^{(1)}X_{i_2j}^{(m)} \right) \right. \\ \hspace{3cm}
	\left. +\sum_{l=2}^N  \sum_{m=1}^N g(\zeta_l(\gamma) \lambda_m ) \left(Y_{jj,r}^{(l)}X_{jj}^{(m)} - Y_{i_1j,r}^{(l)}X_{i_2j}^{(m)} \right) \right]. \label{Tiij_XY_apendix_sums_2}
\end{eqnarray}
Then, by substituting Eq.~(\ref{Yijm_appendix_1}) into Eq. (\ref{Tiij_XY_apendix_sums_2}), we obtain
\begin{eqnarray}
	\nonumber
	\displaystyle
	\fl 
	\langle T(i_1,i_2,r,j) \rangle = \frac{1}{P_j^\infty(r;\gamma) P_j^\infty } \Bigg[ \delta_{i_1j} \delta_{i_2j} \\ \nonumber \fl + \sum_{m=1}^N  g(\zeta_1(\gamma) \lambda_m ) \left( X_{jj}^{(1)} X_{jj}^{(m)}  + \gamma \sum^N_{k=2} \frac{X^{(k)}_{rj}X_{jj}^{(m)}}{1-(1-\gamma)\lambda_k}  - X_{i_1j}^{(1)}X_{i_2j}^{(m)}- \gamma \sum^N_{k=2} \frac{X^{(k)}_{rj}X_{i_2j}^{(m)}}{1-(1-\gamma)\lambda_k} \right)  \\ \fl  \left. + \sum_{l=2}^N \sum_{m=1}^N g(\zeta_l(\gamma) \lambda_m ) \left(X_{jj}^{(l)} X_{jj}^{(m)}-   \frac{\gamma X^{(l)}_{rj} X_{jj}^{(m)}}{1-(1-\gamma)\lambda_l} - X_{i_1j}^{(l)}X_{i_2j}^{(m)} +   \frac{\gamma X^{(l)}_{rj}X_{i_2j}^{(m)}}{1-(1-\gamma)\lambda_l} \right) \right].
\end{eqnarray}
In order to simplify this result, we first rearrange the expression as follows
\begin{eqnarray}
	\fl \nonumber
	\langle T(i_1,i_2,r,j) \rangle = \frac{1}{P_j^\infty(r;\gamma) P_j^\infty } \Bigg[ \delta_{i_1j} \delta_{i_2j}\\  \nonumber \fl + \sum_{m=1}^N g(\zeta_1(\gamma) \lambda_m ) \left( X_{jj}^{(1)} X_{jj}^{(m)} - X_{i_1j}^{(1)}X_{i_2j}^{(m)} + \gamma \sum^N_{l=2} \left(\frac{X^{(l)}_{rj}X_{jj}^{(m)}}{1-(1-\gamma)\lambda_l} - \frac{X^{(l)}_{rj}X_{i_2j}^{(m)}}{1-(1-\gamma)\lambda_l} \right) \right)  \\ \fl \left.  + \sum_{l=2}^N \sum_{m=1}^N g(\zeta_l(\gamma) \lambda_m ) \left(X_{jj}^{(l)} X_{jj}^{(m)}- X_{i_1j}^{(l)}X_{i_2j}^{(m)} -   \frac{\gamma X^{(l)}_{rj} X_{jj}^{(m)}}{1-(1-\gamma)\lambda_l} +   \frac{\gamma X^{(l)}_{rj}X_{i_2j}^{(m)}}{1-(1-\gamma)\lambda_l} \right)  \right].
\end{eqnarray}
Reorganizing the sums, we have
\begin{eqnarray}
\fl \nonumber
	\langle T(i_1,i_2,r,j) \rangle = \frac{1}{P_j^\infty(r;\gamma) P_j^\infty } \left[ \delta_{i_1j} \delta_{i_2j} + \sum_{l,m=1}^N g(\zeta_l(\gamma) \lambda_m ) \left( X_{jj}^{(l)} X_{jj}^{(m)} - X_{i_1j}^{(l)}X_{i_2j}^{(m)} \right) \right. \\ \nonumber \left. \left. \left.  +\gamma   \sum^N_{l=2}\sum_{m=1}^N g(\zeta_1(\gamma) \lambda_m )\left(  \frac{X^{(l)}_{rj}X_{jj}^{(m)}}{1-(1-\gamma)\lambda_l} - \frac{X^{(l)}_{rj}X_{i_2j}^{(m)}}{1-(1-\gamma)\lambda_l} \right) \right. \right. \right.  \\
	\left.  - \gamma \sum_{l=2}^N \sum_{m=1}^N g(\zeta_l(\gamma) \lambda_m ) \left(  \frac{X^{(l)}_{rj} X_{jj}^{(m)}}{1-(1-\gamma)\lambda_l} - \frac{X^{(l)}_{rj}X_{i_2j}^{(m)}}{1-(1-\gamma)\lambda_l} \right) \right]
\end{eqnarray}
and, factorizing $\gamma$, we have
\newpage
\begin{eqnarray}\nonumber
	\fl
	\langle T(i_1,i_2,r,j) \rangle = \frac{1}{P_j^\infty(r;\gamma) P_j^\infty } \left[ \delta_{i_1j} \delta_{i_2j} + \sum_{l,m=1}^N g(\zeta_l(\gamma) \lambda_m ) \left( X_{jj}^{(l)} X_{jj}^{(m)} - X_{i_1j}^{(l)}X_{i_2j}^{(m)} \right) \right. \\  \left.  +\gamma  \sum^N_{l=2} \sum_{m=1}^N \frac{g(\zeta_1(\gamma) \lambda_m )-g(\zeta_l(\gamma) \lambda_m )}{1-(1-\gamma)\lambda_l} \left(  X^{(l)}_{rj}X_{jj}^{(m)} - X^{(l)}_{rj}X_{i_2j}^{(m)}\right) \right]. \label{appendix_sums_45}
\end{eqnarray}
Now, considering the case when $m=1$ in the sums proportional to $\gamma$ in Eq. (\ref{appendix_sums_45}), we obtain a term of the following form
\begin{equation}
	X^{(l)}_{rj}X_{jj}^{(1)} - X^{(l)}_{rj}X_{i_2j}^{(1)}.
\end{equation}
Here, $X_{i_2j}^{(1)} = X_{jj}^{(1)}$ because it represents the same expression for the stationary distribution of a normal walker $P_j^\infty$, which is independent of the starting node. Therefore, for the case when $m=1$, we obtain  $X^{(l)}_{rj}X_{jj}^{(1)} - X^{(l)}_{rj}X_{i_2j}^{(1)}=(X^{(l)}_{rj}- X^{(l)}_{rj})P_j^\infty=0$. Then
 Eq. (\ref{appendix_sums_45}) takes the form
\begin{eqnarray}\nonumber
	\fl
	\langle T(i_1,i_2,r,j) \rangle = \frac{1}{P_j^\infty(r;\gamma) P_j^\infty } \left[ \delta_{i_1j} \delta_{i_2j} + \sum_{l,m=1}^N g(\zeta_l(\gamma) \lambda_m ) \left( X_{jj}^{(l)} X_{jj}^{(m)} - X_{i_1j}^{(l)}X_{i_2j}^{(m)} \right) \right. \\ \left. \left. \left.  +\gamma\sum_{l,m=2}^N \frac{g(\zeta_1(\gamma) \lambda_m )-g(\zeta_l(\gamma) \lambda_m )}{1-(1-\gamma)\lambda_l} \left(  X^{(l)}_{rj}X_{jj}^{(m)} - X^{(l)}_{rj}X_{i_2j}^{(m)}\right) \right. \right.  \right]. \label{appendix_sums_47}
\end{eqnarray}
Finally, the value of $\frac{g(\zeta_1(\gamma) \lambda_m )-g(\zeta_l(\gamma) \lambda_m )}{1-(1-\gamma)\lambda_l}$ needs to be determined. To do this, we explicitly express it by considering
\begin{equation}\label{g_eigen}
	g(z) \equiv \left\{
	\begin{array}{ll}
		\displaystyle
		(1 - z)^{-1} & \qquad \mathrm{for}\ z \neq 1, \\
		0            & \qquad \mathrm{for}\ z = 1.
	\end{array}
	\right.
\end{equation}
We recall that $\zeta_l(\gamma) = (1 - \gamma) \lambda_l$ for $l = 2, \ldots, N$, while $\zeta_1(\gamma) = 1$. We now proceed to develop this expression as follows
\begin{eqnarray}\nonumber
	&\frac{g(\zeta_1(\gamma) \lambda_m )-g(\zeta_l(\gamma) \lambda_m )}{1-(1-\gamma)\lambda_l}= \frac{\frac{1}{1-\lambda_m}-\frac{1}{1-(1-\gamma)\lambda_l\lambda_m}}{1-(1-\gamma)\lambda_l}\\ \nonumber&=\frac{(1-(1-\gamma)\lambda_l\lambda_m)-(1-\lambda_m)}{(1-(1-\gamma)\lambda_l\lambda_m)(1-\lambda_m)(1-(1-\gamma)\lambda_l)}\\  &=\frac{\lambda_m(1-(1-\gamma)\lambda_l)}{(1-(1-\gamma)\lambda_l\lambda_m)(1-\lambda_m)(1-(1-\gamma)\lambda_l)}\nonumber \\  &= \frac{\lambda_m}{(1-\lambda_m)(1-(1-\gamma)\lambda_l\lambda_m)}.
\end{eqnarray}
We can now substitute this value into Eq. (\ref{appendix_sums_47}) to obtain
\begin{eqnarray*}
	\fl
	\langle T(i_1,i_2,r,j) \rangle = \frac{1}{P_j^\infty(r;\gamma) P_j^\infty } \left[ \delta_{i_1j} \delta_{i_2j} + \sum_{l,m=1}^N g(\zeta_l(\gamma) \lambda_m ) \left( X_{jj}^{(l)} X_{jj}^{(m)} - X_{i_1j}^{(l)}X_{i_2j}^{(m)} \right) \right. \\ \left. \left. \left.  +\gamma\sum_{l,m=2}^N \frac{\lambda_m}{(1-\lambda_m)(1-(1-\gamma)\lambda_l\lambda_m)} \left(  X^{(l)}_{rj}X_{jj}^{(m)} - X^{(l)}_{rj}X_{i_2j}^{(m)}\right) \right. \right.  \right].
\end{eqnarray*}
This result is presented in the main text in Eq.~(\ref{T_encounter_reset1_normal2}) and provides the mean first-encounter times of a resetting walker and a normal walker in terms of the eigenvalues and eigenvectors of the transition matrix that defines the normal random walker. 
\subsection{Derivation of Eq.~(\ref{Tiijr_2reinicio})}
\label{Ref_Sec_Append_2}
In this appendix, we present the derivation leading to Eq. (\ref{Tiijr_2reinicio}), starting from Eq. (\ref{Tiij_YY}). Our objective is to obtain an analytical expression for the mean first-encounter time of two random walkers, both subjected to resetting dynamics. For clarity, we denote the resetting parameter, resetting node, and initial node of the first walker with subscript $1$, and those of the second walker with subscript $2$. Additionally, we introduce the compact notation $\vec{i} = (i_1, i_2)$ and $\vec{r} = (r_1, r_2)$. The derivation begins with Eq.~(\ref{Tiij_YY})
\begin{equation}\label{Tiij_YY_apendix}
	\fl
	\langle T(\vec{i},\vec{r},j) \rangle = \frac{1}{P_{j,r_1}^\infty P_{j,r_2}^\infty} \left[ \delta_{i_1j} \delta_{i_2j} + \sum_{l,m=1}^N g(\zeta_l(\gamma_1) \zeta_m(\gamma_2) ) \left(Y_{jj,r_1}^{(l)}Y_{jj,r_2}^{(m)} - Y_{i_1j,r}^{(l)}Y_{i_2j,r_2}^{(m)} \right) \right],
\end{equation}
where we use the more compact notation $P_{j,r}^\infty$ to denote $P_j^\infty(r;\gamma)$. In addition,  $Y_{ij,r}^{(l)}$ is defined in Eq. (\ref{Yijm_appendix_1}).
\\[2mm]
We identify three relevant cases in the double summation in Eq. (\ref{Tiij_YY_apendix}): (i) $m = 1$ and $l = 2, \ldots, N$; (ii) $m = 2, \ldots, N$ and $l = 1$; and (iii) $m = 2, \ldots, N$ and $l = 2, \ldots, N$. The term corresponding to $m = 1$ and $l = 1$ does not contribute to the sum, since $g(\zeta_1(\gamma_1) \zeta_1(\gamma_2)) = 0$, as the largest eigenvalue is always $\zeta_1 = 1$. Accordingly, the summation is decomposed into the three aforementioned cases as follows
\begin{eqnarray}
	\fl \nonumber
	\langle T(\vec{i},\vec{r},j) \rangle = \frac{1}{P_{j,r_1}^\infty P_{j,r_2}^\infty} \left[ \delta_{i_1j} \delta_{i_2j} + \sum_{l=2}^N  g(\zeta_l(\gamma_1) \zeta_1(\gamma_2) ) \left(Y_{jj,r_1}^{(l)}Y_{jj,r_2}^{(1)} - Y_{i_1j,r}^{(l)}Y_{i_2j,r_2}^{(1)} \right) \right.  \\ \nonumber
	\quad \quad \quad \quad \quad \quad \quad + \sum_{m=2}^N  g(\zeta_1(\gamma_1) \zeta_m(\gamma_2) ) \left(Y_{jj,r_1}^{(1)}Y_{jj,r_2}^{(m)} - Y_{i_1j,r}^{(1)}Y_{i_2j,r_2}^{(m)} \right)  \\ \hspace{2cm}
	\left. + \sum_{l,m=2}^N  g(\zeta_l(\gamma_1) \zeta_m(\gamma_2) ) \left(Y_{jj,r_1}^{(l)}Y_{jj,r_2}^{(m)} - Y_{i_1j,r}^{(l)}Y_{i_2j,r_2}^{(m)} \right) \right]. \label{two_reset_T_sums}
\end{eqnarray}
We analyze each case individually, starting with the case $l = 1$ and $m = 2, \ldots, N$. We substitute according to Eq.~(\ref{Yijm_appendix_1})
\begin{eqnarray}
	\fl
	\nonumber
	Y_{jj,r_1}^{(1)}Y_{jj,r_2}^{(m)} - Y_{i_1j,r}^{(1)}Y_{i_2j,r_2}^{(m)} = X_{jj}^{(1)}X_{jj}^{(m)} - \gamma_2  \frac{X_{jj}^{(1)}X^{(m)}_{r_2j}}{1-(1-\gamma_2)\lambda_m} + \gamma_1 \sum^N_{k=2} \frac{X^{(k)}_{r_1j}X_{jj}^{(m)}}{1-(1-\gamma_1)\lambda_k} \\ 
	\fl
	\nonumber \qquad
	- \gamma_1 \gamma_2 \sum^N_{k=2} \frac{X^{(k)}_{r_1j}X^{(m)}_{r_2j}}{(1-(1-\gamma_1)\lambda_k)(1-(1-\gamma_2)\lambda_m)} - X_{i_1j}^{(1)}X_{i_2j}^{(m)} + \gamma_2  \frac{X_{i_1j}^{(1)}X^{(m)}_{r_2j}}{1-(1-\gamma_2)\lambda_m}\\ - \gamma_1 \sum^N_{k=2} \frac{X^{(k)}_{r_1j}X_{i_2j}^{(m)}}{1-(1-\gamma_1)\lambda_k} + \gamma_1 \gamma_2 \sum^N_{k=2} \frac{X^{(k)}_{r_1j}X^{(m)}_{r_2j}}{(1-(1-\gamma_1)\lambda_k)(1-(1-\gamma_2)\lambda_m)}.
\end{eqnarray}
In this result, the terms proportional to $\gamma_1\gamma_2$ cancels out, and the remaining terms can be grouped as follows
\begin{eqnarray}
\fl 
\nonumber	
\qquad 	Y_{jj,r_1}^{(1)}Y_{jj,r_2}^{(m)} - Y_{i_1j,r}^{(1)}Y_{i_2j,r_2}^{(m)} = \left(X_{jj}^{(1)}X_{jj}^{(m)} - X_{i_1j}^{(1)}X_{i_2j}^{(m)}\right)   \\ \qquad + \gamma_1 \sum^N_{k=2} \frac{\left(X^{(k)}_{r_1j}X_{jj}^{(m)}-X^{(k)}_{r_1j}X_{i_2j}^{(m)}\right)}{1-(1-\gamma_1)\lambda_k} - \gamma_2  \frac{\left(X_{jj}^{(1)}X^{(m)}_{r_2j}-X_{i_1j}^{(1)}X^{(m)}_{r_2j}\right)}{1-(1-\gamma_2)\lambda_m}. \label{appendix_eq_53}
\end{eqnarray}
Considering that $X_{ij}^{(1)}$ corresponds to the stationary distribution of a normal random walker, it only depends on the index $j$. Therefore, in Eq. (\ref{appendix_eq_53}) the term proportional to $\gamma_2$ is $X_{jj}^{(1)} X_{r_2j}^{(m)} - X_{i_1j}^{(1)} X_{r_2j}^{(m)} = P_j^\infty(X_{r_2j}^{(m)} - X_{r_2j}^{(m)})=0$. Consequently, we have
\begin{equation}\label{two_reset_sum_1}
	\fl \,\,
	Y_{jj,r_1}^{(1)}Y_{jj,r_2}^{(m)} - Y_{i_1j,r}^{(1)}Y_{i_2j,r_2}^{(m)} = \left(X_{jj}^{(1)}X_{jj}^{(m)} - X_{i_1j}^{(1)}X_{i_2j}^{(m)}\right)  + \gamma_1 \sum^N_{k=2} \frac{\left(X^{(k)}_{r_1j}X_{jj}^{(m)}-X^{(k)}_{r_1j}X_{i_2j}^{(m)}\right)}{1-(1-\gamma_1)\lambda_k}.
\end{equation}
A similar expression can be derived for the case with $l = 2, \ldots, N$ and $m = 1$
\begin{equation}\label{two_reset_sum_2}
	\fl \,\,
	Y_{jj,r_1}^{(l)}Y_{jj,r_2}^{(1)} - Y_{i_1j,r}^{(l)}Y_{i_2j,r_2}^{(1)} = \left(X_{jj}^{(l)}X_{jj}^{(1)} - X_{i_1j}^{(l)}X_{i_2j}^{(1)}\right) + \gamma_2 \sum^N_{k=2} \frac{\left(X^{(l)}_{jj}X_{r_2j}^{(k)}-X^{(l)}_{i_1j}X_{r_2j}^{(k)}\right)}{1-(1-\gamma_2)\lambda_k},
\end{equation}
and, finally, for the case $l=2,\ldots,N$ and $m=2,\ldots,N$, we obtain
\begin{eqnarray}\nonumber
	\fl 
	\qquad 
	Y_{jj,r_1}^{(l)}Y_{jj,r_2}^{(m)} - Y_{i_1j,r}^{(l)}Y_{i_2j,r_2}^{(m)} = \left(X_{jj}^{(l)}X_{jj}^{(m)} - X_{i_1j}^{(l)}X_{i_2j}^{(m)}\right) \\ \qquad - \gamma_1  \frac{\left(X_{r_1j}^{(l)}X^{(m)}_{jj}-X_{r_1j}^{(l)}X^{(m)}_{i_2j}\right)}{1-(1-\gamma_1)\lambda_l}  - \gamma_2  \frac{\left(X_{jj}^{(l)}X^{(m)}_{r_2j}-X_{i_1j}^{(l)}X^{(m)}_{r_2j}\right)}{1-(1-\gamma_2)\lambda_m}. \label{two_reset_sum_3}
\end{eqnarray}
From the results in Eqs.~(\ref{two_reset_sum_1})-(\ref{two_reset_sum_3}), common terms in the three cases analyzed can be identified and grouped together. Consequently, Eq. (\ref{two_reset_T_sums}) simplifies to the following form
\begin{eqnarray}
	\fl
	\nonumber 
	\langle T(\vec{i},\vec{r},j) \rangle = \frac{1}{P_{j,r_1}^\infty P_{j,r_2}^\infty} \left[ \delta_{i_1j} \delta_{i_2j} + \sum_{l,m=1}^N  g(\zeta_l(\gamma_1) \zeta_m(\gamma_2) ) \left(X_{jj}^{(l)}X_{jj}^{(m)} - X_{i_1j}^{(l)}X_{i_2j}^{(m)} \right) \right.  \\ \fl \nonumber \hspace{2cm} +  \gamma_1 \sum_{l,m=2}^N \frac{ g(\zeta_1(\gamma_1) \zeta_m(\gamma_2) )-g(\zeta_l(\gamma_1) \zeta_m(\gamma_2) )}{1-(1-\gamma_1)\lambda_l}\left(X^{(l)}_{r_1j}X_{jj}^{(m)}-X^{(l)}_{r_1j}X_{i_2j}^{(m)} \right)  \\  \fl  \hspace{2cm} \left. +  \gamma_2 \sum_{l,m=2}^N \frac{ g(\zeta_l(\gamma_1) \zeta_1(\gamma_2) )-g(\zeta_l(\gamma_1) \zeta_m(\gamma_2) ) }{1-(1-\gamma_2)\lambda_m}\left(X^{(l)}_{jj}X_{r_2j}^{(m)}-X^{(l)}_{i_1j}X_{r_2j}^{(m)} \right)\right]. \label{two_reset_T_sums_factors}
\end{eqnarray}
Considering that $\zeta_1(\gamma)$ is the largest eigenvalue and thus equals $1$, it is necessary to evaluate the following expressions
\begin{equation}
\frac{ g(\zeta_m(\gamma_2)) - g(\zeta_l(\gamma_1) \zeta_m(\gamma_2)) }{1 - (1 - \gamma_1) \lambda_l}
\quad \mathrm{and} \quad
\frac{ g(\zeta_l(\gamma_1)) - g(\zeta_l(\gamma_1) \zeta_m(\gamma_2)) }{1 - (1 - \gamma_2) \lambda_m}.
\end{equation}
Following a similar approach to the one used in the case analyzed in Appendix \ref{Ref_Sec_Append_1}, it is found that these expressions simplify to
\begin{equation}\nonumber
	\fl
	\qquad
	\frac{ g(\zeta_m(\gamma_2) )-g(\zeta_l(\gamma_1) \zeta_m(\gamma_2) )}{1-(1-\gamma_1)\lambda_l}= \frac{\lambda_m(1-\gamma_2)}{(1-(1-\gamma_2)\lambda_m)(1-(1-\gamma_1)(1-\gamma_2)\lambda_l\lambda_m)},
\end{equation}
\begin{equation}\nonumber
	\fl
	\qquad
	\frac{g( \zeta_l(\gamma_1) )-g(\zeta_l(\gamma_1) \zeta_m(\gamma_2) )}{1-(1-\gamma_2)\lambda_m}= \frac{\lambda_l(1-\gamma_1)}{(1-(1-\gamma_1)\lambda_l)(1-(1-\gamma_1)(1-\gamma_2)\lambda_l\lambda_m)}.
\end{equation}
This result, incorporated in Eq.~(\ref{two_reset_T_sums_factors}), allows obtaining the mean first-encounter times
\begin{eqnarray*}
	\fl
	\langle T(\vec{i},\vec{r},j) \rangle = \frac{1}{P_{j,r_1}^\infty P_{j,r_2}^\infty} \left[ \delta_{i_1j} \delta_{i_2j} + \sum_{l,m=1}^N  g(\zeta_l(\gamma_1) \zeta_m(\gamma_2) ) \left(X_{jj}^{(l)}X_{jj}^{(m)} - X_{i_1j}^{(l)}X_{i_2j}^{(m)} \right) \right.  \\ 
	\fl \quad \quad \quad  +  \gamma_1 \sum_{l,m=2}^N  \frac{\lambda_m(1-\gamma_2)}{(1-(1-\gamma_2)\lambda_m)(1-(1-\gamma_1)(1-\gamma_2)\lambda_l\lambda_m)}\left(X^{(l)}_{r_1j}X_{jj}^{(m)}-X^{(l)}_{r_1j}X_{i_2j}^{(m)} \right)  \\ \fl \quad \qquad  \left. +  \gamma_2 \sum_{l,m=2}^N \frac{\lambda_l(1-\gamma_1)}{(1-(1-\gamma_1)\lambda_l)(1-(1-\gamma_1)(1-\gamma_2)\lambda_l\lambda_m)}\left(X^{(l)}_{jj}X_{r_2j}^{(m)}-X^{(l)}_{i_1j}X_{r_2j}^{(m)} \right)\right].
\end{eqnarray*}
This equation allows us to obtain the mean first-encounter times of two simultaneous random walkers with resetting in terms of the eigenvalues and eigenvectors of a normal random walker, as presented in the main text in Eq.~(\ref{Tiijr_2reinicio}).

\section*{Acknowledgments}
D.R.G. acknowledges support from CONAHCYT Mexico.

\section*{References}

\providecommand{\newblock}{}

\end{document}